\documentclass[
reprint,
aip, jap,
superscriptaddress,
floatfix
]{revtex4-1}

\usepackage{makeidx}

\usepackage{graphicx}
\usepackage{amsmath}
\usepackage{epstopdf}
\usepackage{textcomp}
\usepackage{epstopdf}
\usepackage{tabularx}
\usepackage{ulem} 
\usepackage{color}

\usepackage{natbib}
\usepackage{bm}%
\usepackage[colorlinks=true,linkcolor=blue]{hyperref}%
\expandafter\ifx\csname package@font\endcsname\relax\else
 \expandafter\expandafter
 \expandafter\usepackage
 \expandafter\expandafter
 \expandafter{\csname package@font\endcsname}%
\fi
\makeindex

\begin{document}

\title{\color{blue} Spin-pumping through a varying-thickness MgO interlayer in Fe/Pt system}

\author{Laura Mihalceanu}
\author{Sascha Keller}
\author{Jochen Greser}

\affiliation{Fachbereich Physik and Landesforschungszentrum OPTIMAS, Technische Universit\"{a}t Kaiserslautern,
Erwin-Schr\"{o}dinger-Str. 56, 67663 Kaiserslautern, Germany}

\author{Dimitrios Karfaridis}
\author{Konstantinos Symeonidis}
\author{Georgios Vourlias}
\author{Thomas Kehagias}

\affiliation{Physics Department, Aristotle University of Thessaloniki, 54124 Thessaloniki, Greece}

\author{Andres Conca}

\author{Burkard Hillebrands}

\author{Evangelos Th. Papaioannou}
\thanks{Author to whom correspondence should be addressed. Email: epapa@physik.uni-kl.de}

\affiliation{Fachbereich Physik and Landesforschungszentrum OPTIMAS, Technische Universit\"{a}t Kaiserslautern,
Erwin-Schr\"{o}dinger-Str. 56, 67663 Kaiserslautern, Germany}


\begin{abstract}

The spin-pumping mechanism is probed through a tunnelling MgO interlayer in  Fe/Pt bilayers. We show by ferromagnetic resonance technique and spin-pumping experiments that spin currents can tunnel through the MgO interlayer for thickness up to 2~{nm} and can produce significant voltages in the Pt layer. The electrical detection of spin-pumping furthermore reveals the critical role of rectification and shunting effects on the generated voltages. The  non zero spin current transport through a few monolayers of an insulating interlayer might initiate further studies on the role of very thin oxides in spin-pumping experiments.

\end{abstract}



\maketitle {}


The generation of a spin current via the spin-pumping effect (SP) and its detection through the inverse spin Hall effect (ISHE) is a key topic of research in the spintronics community.~\cite{6516040, RevModPhys.87.1213} A spin current is generated at the interface between a ferromagnetic material (FM) or a ferrimagnetic insulator (for example YIG) and a non magnetic metal (NM) by a precessing magnetization in the magnetic layer. A large part of the research on spin-pumping has been focused on the YIG/Pt system,~\cite{PhysRevLett.111.106601,Jungfleisch-thickness} a system that has also enabled the discovery of novel effects like the spin Hall magnetoresistance.~\cite{PhysRevB.87.144411,PhysRevLett.110.206601,PhysRevB.87.184421} As well, metallic bilayers like the Py/Pt system~\cite{Ando2011,vilela-leao:07C910} have attracted a lot of interest. The efficiency of spin-pumping depends on the transparency of the interface quantified by the parameter termed spin mixing conductance~\cite{Tserkovnyak2002} and is investigated in a wide range of bilayer systems.~\cite{PhysRevLett.111.176601,PhysRevLett.107.046601,Papaioannou2013,Parkin2015} The critical role of the interface transparency accounts for the ability to transfer angular momentum at the FM-NM interface. It is widely accepted that the transfer depends on the exchange interaction between the precessing magnetization (of the FM layer) and the conduction electrons (of the NM layer).~\cite{6516040} Here, we structurally modify the interface transparency by implementing a tunnelling MgO barrier at the FM/NM interface in order to probe the spin pumping effect. 

Previous research  on the impact of an insulating interlayer in spin-pumping experiments has yielded contradictory conclusions. In the study of Moriyama et al.~\cite{Moriyama2008} on Al/AlO$_{x}$/Ni$_{80}$Fe$_{20}$/Cu tunnel junctions a large voltage was detected at  spin-pumping conditions, however  a quantitative analysis showed a number of discrepancies with the standard spin-pumping theory. In the works of Mosendz et.al.~\cite{Mosendz2010} on Py/MgO/Pt   and Kim et al.~\cite{Kim2011} on Py/nanooxide/Pd the  spin-pumping induced voltage was suppressed by the tunneling barrier structures. Also, in these studies no systematic analysis on the thickness dependence of the insulating layer was performed. Recently,  the spin-pumping mechanism was further examined in YIG/insulating barrier (Sr$_{2}$GaTaO$_{6}$,SrTiO$_{3}$, Sr$_{2}$CrNbO$_{6}$)/Pt structures.~\cite{PhysRevLett.111.247202} An exponential decay of the measured voltage originating from the spin-pumping was concluded by varying the thickness of the barrier. 

In this work, a thickness dependent study of the influence of an insulating tunnelling barrier layer in a metallic system is performed. As a metallic bilayer host for the insertion of the barrier layer we use a fully epitaxial Fe/Pt bilayer system. We choose Fe/Pt as a model system where the interface engineering has already shown the influence on the efficiency of spin-pumping.~\cite{Conca2016,Papaioannou2013} As a barrier material we use MgO  with the aim not only to investigate the role of tunneling properties of the conduction electrons at the Fe/MgO/Pt interfaces but also to probe the flow of spin current through MgO/Fe interface well known for its pronounced spin-dependent tunnelling  properties in Fe/MgO/Fe tunnel junctions. We show that the presence of a very thin MgO barrier of up to 2~{nm} thickness only partially prevents the flow of a spin current.

Molecular beam epitaxy (MBE) was used to grow a series of Fe/MgO/Pt samples with MgO thicknesses ranging from 0.5~{nm} to 2~{nm}. The thickness of the Fe layer of 12~{nm} and of the Pt layer of 6~{nm} have been kept constant. Two more films were fabricated as reference samples: Fe~(12~{nm})~/~Pt (6~{nm})  and Fe (12~{nm})~/~MgO~(10~{nm}). For the latter the MgO overlayer functions as a protection against oxidation.

\begin{figure}
 \includegraphics[width =1.0 \columnwidth]{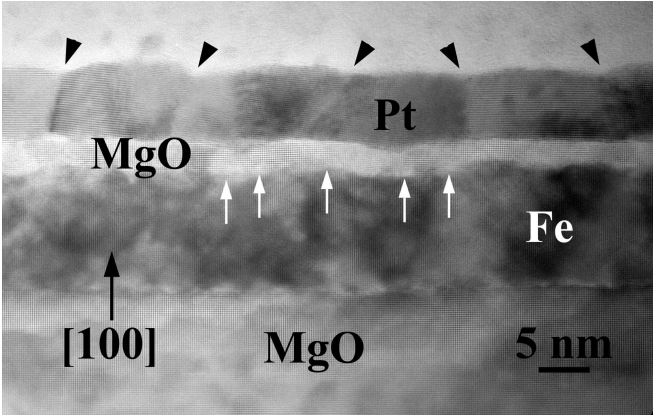}
\caption{\label{fig:tem} HRTEM image, along the [001] MgO projection, illustrating the atomic structure of the multilayer. Fe is oriented along the [1-10] projection direction. The thin MgO interlayer is relaxed through the introduction of an irregular array of misfit dislocations at the Fe/MgO interface (white arrows). Furthermore, the Pt layer is polycrystalline consisting of misoriented crystals almost equal in thickness, which are joined by low-angle grain boundaries (black triangles).}
\end{figure}

X-ray diffraction (XRD) measurements established the epitaxial relationship and the crystallographic orientation of the Fe/MgO/Pt films. The growth of Fe on (100) MgO is featured by a $45^{\circ}$  in-plane rotation with respect to the substrate in all samples, owing to the \textit{fcc}/\textit{bcc} unit cell stratification. The XRD patterns of the Pt overlayer reveal the formation of a textured growth oriented along the [111] direction. However, by increasing the MgO thickness above 1~{nm}, the appearance of (200) peaks together with the (111) peak is recorded, pointing out that  Pt crystals with [100] growth direction are also present. It was unfeasible for X-rays to disclose qualitative data from the insulated MgO interlayer, due to the dominant contribution of the substrate. Therefore, high resolution transmission electron microscopy (HRTEM) was employed to analyze the local structure of the multilayer. In Fig.~\ref{fig:tem}  the atomic structure of the sample comprising a 2~{nm} thick MgO interlayer along the [001] zone axis of MgO is shown. It reveals the epitaxial growth of MgO/Fe/MgO. In particular, in the Fe layer only the (110) lattice fringes are resolved, due to the $45^{\circ}$ in-plane rotation of the \textit{bcc} Fe lattice relative to both the MgO substrate and interlayer \textit{fcc} lattices. The Fe layer exhibits an $\mathrm{r_{RMS} \simeq 0.5~nm}$ surface roughness which the  overlying MgO layer follows. The MgO interlayer is continuous, it grows epitaxially on Fe and appears to be stress-free, due to the introduction of an array of misfit dislocations at the Fe/MgO interface. Their projected edge components are denoted by white arrows in Fig.~\ref{fig:tem}. Indeed, direct measurements on HRTEM images resulted in lattice plane spacing values of $\mathrm{d_{110}(Fe)=0.203~nm}$ and $\mathrm{d_{020}(MgO)=0.210~nm}$, consistent with a relaxed configuration. HRTEM confirmed the textured growth of the Pt layer that comprises of crystals of 5-15~{nm} lateral dimensions. They are connected by low-angle grain boundaries, indicated by the black triangles in Fig.~\ref{fig:tem}. In addition, HRTEM imaging and electron diffraction local analysis detected  Pt crystals grown along the [100] direction in the observable area of the sample. A substantial number of Pt crystals were oriented off the [001] Pt viewing axis exhibiting arbitrary in-plane rotations, while retaining the [100] growth orientation within a $10^{\circ}$ tilt angle around the normal to the substrate direction. 

\begin{figure}
\includegraphics[width =1.0 \columnwidth]{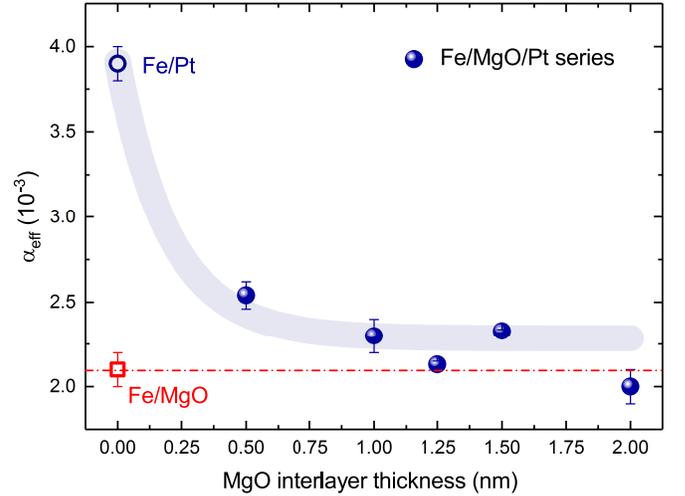}
\caption{\label{fig:alpha} Effective Gilbert damping parameter $\alpha_{\mathrm{eff}}$ with respect to the MgO interlayer thickness. $\alpha_{\mathrm{eff}}$ is decreasing from the Fe/Pt value with increasing MgO thickness and approaches the value of the Fe reference film for the 2~{nm} MgO interlayer. The thick line serves as a guide to the eye.}
\end{figure}

Initially, we have addressed the spin-pumping effect by measuring the additional contribution to the effective Gilbert damping. The study of the dynamic magnetization behavior has been examined by performing ferromagnetic resonance (FMR) measurements along the Fe easy axis.  In a FMR experiment, the ferromagnetic resonance is excited by a strip-line antenna and the resonance signal is measured using a vector network analyzer (FMR-VNA). All the samples were placed with the Pt side facing the strip line antenna and the static magnetic field \textit{H} is aligned parallel to the antenna. The $S_{12}$ transmission parameter is recorded.  By fitting the dependence of the linewidth $\Delta H$ on the resonance frequency \cite{Conca2016}  the effective Gilbert damping parameter $\alpha_{\mathrm{eff}}$ is obtained, and is shown in Fig.~\ref{fig:alpha}. Due to the absence of a spin sink, the Fe/MgO reference  will not generate a spin current by means of spin-pumping and, as expected, exhibits the lowest effective Gilbert damping value, $\alpha_\text{Fe/MgO}$. On the other side,  Fe/Pt possesses the highest value of $\alpha_{\mathrm{eff}}$ of almost twice the value of the Fe/MgO control sample. This is due to the generation of a spin current through the Fe/Pt interface by means of the spin-pumping effect. All samples with an MgO interlayer have lower $\alpha_{\mathrm{eff}}$ values compared to the Fe/Pt. By gradually increasing the thickness of the MgO interlayer, $\alpha_{\mathrm{eff}}$ is decreasing, indicating a reduction of the spin current injection until being completely blocked at an MgO thickness of $2\mathrm{~nm}$.

To further quantify the  spin current generation due to the spin-pumping effect the effective spin-mixing conductance \cite{RevModPhys.77.1375,Conca2016,Carva2007} parameter was evaluated using:

\begin{equation}
g^{\mathrm{\uparrow\downarrow}}_{\mathrm{eff}} = \frac{ 4 \pi{M_\text{{s}} d_\text{Fe}}}{\gamma \mu_\text{B}} (\alpha_{\mathrm{eff}} -\alpha_\text{Fe/MgO}).	
\label{fig:mixing}
\end{equation} 

\noindent where \textit{d}$_\text{Fe}$ is the thickness of the Fe layer that is kept constant throughout the sample series and \textit{M}$_\text{s}$ is the saturation magnetization. In our calculation we have used for \textit{M}$_\text{s}$ the effective saturation magnetization obtained from the FMR measurements which is found to be in the range of 1730-1760 kA/m.

Equation~\ref{fig:mixing} provides $g^\mathrm{\uparrow\downarrow}_{\mathrm{eff}}$ values that are decreasing with the thickness of the MgO interlayer. In particular, $g^\mathrm{\uparrow\downarrow}_\mathrm{Fe/Pt} = (2.59 \pm 0.15 )\cdotp 10^{19} \mathrm{~m^{-2}}$, $g^\mathrm{\uparrow\downarrow}_\mathrm{0.5 \mathrm{nm} MgO} = (0.62 \pm 0.19) \cdotp 10^{19} \mathrm{~m^{-2}}$, and $g^\mathrm{\uparrow\downarrow}_\mathrm{1 \mathrm{nm} MgO} = (0.33 \pm 0.06) \cdotp 10^{19} \mathrm{~m^{-2}}$ for the Fe/Pt reference, MgO $0.5~\mathrm{nm}$, and MgO $1~\mathrm{nm}$ respectively.


\begin{figure}
\includegraphics[width =1.0 \columnwidth]{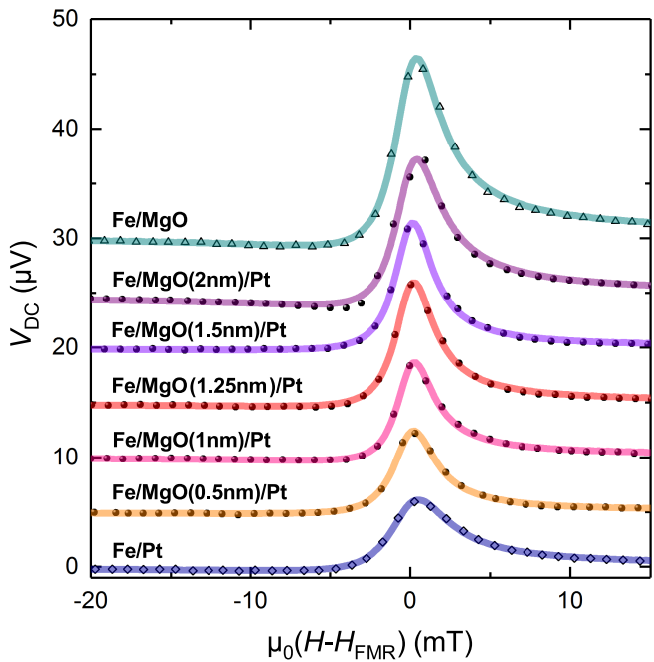}
\caption{Dependency of DC voltages on the external applied magnetic field \textit{H} normalized by the resonant magnetic field \textit{H}$_\text{FMR}$. The curves have been shifted vertically with a 5 $\mu$V step for clarity. The highest voltage is measured for the Fe/MgO reference and the lowest for the Fe/Pt reference sample. The trilayers have voltages that are increasing with the MgO interlayer thickness. The lines are fits according to Eq.~\ref{ISHE fit}.}
\label{fig:ishe-measur}
\end{figure}

The spin-pumping induced ISHE in Pt was probed by measuring the transverse (to the antenna) DC voltage, $V_{\mathrm{DC}}$, by a lock-in amplification technique.~\cite{Papaioannou2013}  The measurement geometry for the spin-pumping experiments was chosen analogous to the FMR experiments. The measurements have been performed at an excitation frequency of 13~{GHz} with a microwave applied power of  24~{dBm}.

In Fig.~\ref{fig:ishe-measur}  the  dependencies of the $V_{\mathrm{DC}}$  voltages on the external applied magnetic field \textit{H} normalized to the resonant magnetic field \textit{H}$_\text{FMR}$ are presented. The curves have been vertically shifted for better visibility. Voltage peaks are seen at external magnetic fields matching the ferromagnetic resonance condition of the magnetic layer. For the Fe/MgO/Pt trilayers  the highest absolute $V_{\mathrm{DC}}$  generation is revealed by the 2~{nm} thick MgO interlayer sample, for which according to $\alpha_{\mathrm{eff}}$ the lowest spin current transport is expected (see Fig.~\ref{fig:alpha}). The Fe reference film capped with an MgO protection layer shows an even higher $V_{\mathrm{DC}}$ while the lowest $V_{\mathrm{DC}}$ is observed for the Fe/Pt system. In order to understand this behaviour  the presence of spin rectification effects which can add up to the pure ISHE voltage must be considered, as well as the shunting effect caused by the capping layer. Spin rectification effects refer to the generation of a static voltage due to the coupling of the high frequency magnetization dynamics near the FMR condition to an rf current.~\cite{Juretschke1960} When the overlayer is Pt, the measured voltage is lower since Pt shunts the voltage generated by the rectification effects.~\cite{Saitoh2016} When the overlayer is MgO shunting is prohibited and the electrical detection of the FMR of the Fe film due to rectification provides the highest $V_{\mathrm{DC}}$. The samples with the MgO interlayer exhibit $V_{\mathrm{DC}}$ values in between the two reference samples.Thicker interlayers reduce the shunting proportionally. Moreover, $V_{\mathrm{DC}}$ of the trilayers depends on the excitation frequency. Measurements in a range of 11-15 {GHz} showed a shift of the $V_{\mathrm{DC}}$  peak towards higher resonant fields for higher frequencies (as expected for the FMR condition), as well as a decrease of  the $V_{\mathrm{DC}}$ values. The latter indicates that each  trilayer behaves like a capacitor with capacitive resistance \textit{R}$_{\mathrm{C}}= 1/\omega C$. The capacitance $C$ depends on the thickness of the dielectric (MgO) in between the two metallic plates (Fe, Pt layers). $C$ is decreasing for higher MgO thicknesses, so the capacitive resistance is expected to increase for the same excitation frequency for thicker MgO interlayers leading to higher $V_{\mathrm{DC}}$ (Fig.~\ref{fig:ishe-measur}).

\begin{figure}
\includegraphics[width =1.0 \columnwidth]{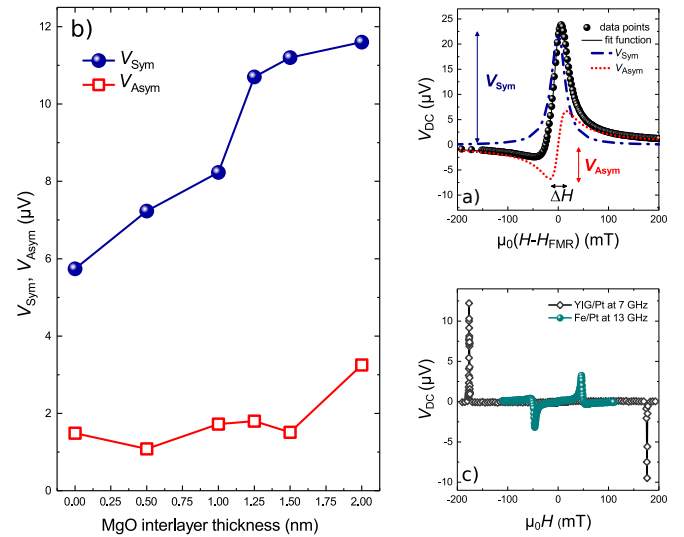}
\caption{(a) Line shape analysis method for the separation of symmetric and antisymmetric voltage contribution. $V_{\mathrm{Sym}}$ and $V_{\mathrm{Asym}}$ are the symmetric and antisymmetric voltage amplitudes, $\Delta H$ is the linewidth and $H_{\text{FMR}}$ the resonance field. (b)  $V_{\mathrm{Sym}}$ and  $V_{\mathrm{Asym}}$ voltage dependence on the MgO thickness, extracted from the fits in Figure~\ref{fig:ishe-measur}. The reference Fe/Pt system is represented at MgO = 0 thickness. Smaller  $V_{\mathrm{Sym}}$ values are indicative of the flow of spin currents through very thin MgO interlayers. (c) The comparison of spin pumping experiments for YIG/Pt and Fe/Pt bilayers performed under the same experimental conditions proves the sign reversal of the generated voltages.
}
\label{fig:ishe_analysis}
\end{figure}

The $V_{\mathrm{DC}}$  generated under spin pumping conditions can be described by charge based effects that arise from the  presence of spin rectification, as aforementioned. On the other side, the non-zero $\alpha_{\mathrm{eff}}$  values in Fig.~\ref{fig:alpha}  indicate the presence of ISHE which generates an additional voltage contribution in the electrically detected signal. The ISHE voltage originated by spin-pumping is known to be purely symmetric.~\cite{Saitoh2016} Figure~\ref{fig:ishe-measur} shows that the detected voltages have symmetric as well as antisymmetric Lorentzian contributions. Spin rectification contributes to the symmetric as well to the antisymmetric $V_{\mathrm{DC}}$ amplitudes. In order  to provide an estimation of the relation of  spin-pumping to rectification effects the  symmetric and antisymmetric components of $V_{\mathrm{DC}}$ were separated by performing a line-shape analysis.~\cite{Azevedo2011}  Following this method which is schematically presented in Fig.~\ref{fig:ishe_analysis} (a), the data are fitted by~\cite{Papaioannou2013}:
\begin{equation}\label{ISHE fit}
\begin{split}
V_{\text{DC}}(H) = & V_{\text{off}} + V_{\mathrm{Sym}} \frac {(\Delta H)^{2} }{(H-H_{\text{FMR}}) ^{2} + {(\Delta H)^{2} }}\\ 
&\quad + V_{\mathrm{Asym}} \frac { -2\Delta H(H-H_{\text{FMR}}) } {(H-H_{\text{FMR}}) ^{2} + (\Delta H)^{2} },
\end{split}
\end{equation}
\noindent where $V_{\mathrm{Sym}}$ and $V_{\mathrm{Asym}}$ are the symmetric and antisymmetric $V_{\mathrm{DC}}$ amplitudes, $V_{\text{off}}$ is a constant offset voltage, $\Delta H$ is the linewidth and $H_{\text{FMR}}$ the magnetic resonance field.$V_{\mathrm{Sym}}$ and $V_{\mathrm{Asym}}$ are plotted in Fig.~\ref{fig:ishe_analysis}~(b). 
The antisymmetric signal appears to be slightly enhanced by the increase of the MgO thickness pointing out  that already the 1~{nm} thick MgO layer is capable to reduce the  shunting that Pt causes. The symmetric signal is also increasing with the MgO thickness for the same reason, however, the increase is more pronounced with respect to the antisymmetric values. The reason for this is the presence of an additional voltage contribution due to the  ISHE. The symmetric signal is a result of two overlapping voltages due to ISHE and rectification. The evolution of the signal can be justified only when the overlapping voltages exhibit opposite sign.

In particular, the smallest symmetric voltage is obtained  for the largest ISHE voltage  that is generated in the Fe/Pt reference sample. The spin-pumping in Fe/Pt causes a significant ISHE voltage due to the presence of an unblocked spin current. As the MgO interlayer is introduced the ISHE voltage contribution is reduced and that results in an increase of the symmetric value. Taking this into account the apparent  increase of the symmetric voltage in Fig.~\ref{fig:ishe_analysis}~(b)  for thicker MgO interlayer thicknesses can be justified as follows: the ISHE generated voltage contribution is subtracted from a symmetric contribution due to rectification. For the 2 nm thick MgO the ISHE voltage contribution is minimized, so  the symmetric component due to rectification approaches a maximum value. 

The different signs between ISHE- and rectification generated voltages are confirmed in Fig.~\ref{fig:ishe_analysis} (c). There, we  compare under the same experimental spin-pumping conditions YIG/Pt (where negligible rectification effects are expected) and Fe/Pt bilayers. We see that for the same direction of the applied magnetic field the sign of the signal is reversed for the metallic Fe/Pt system where the presence of rectification effect is also expected. Since the sign of the spin Hall angle in Pt is the same in both measurements the sign of the ISHE voltage should be the same for Fe/Pt and YIG/Pt for the same orientation of the magnetic field. The reversal in sign can be explained by the presence of rectification effects in Fe/Pt.  Equivalent results have been obtained for Py/Pt~\cite{Sascha2017} and CoFeB/Pt~\cite{Conca2017} systems. 

In summary, we have studied the flow of spin current through an MgO/Fe interface well known for its pronounced spin-dependent tunnelling  properties. The measurement of $\alpha_{\mathrm{eff}}$ with ferromagnetic resonance experiments have revealed that spin-pumping is taking place for MgO thicknesses $< 2$~{nm}. The presence of an MgO interlayer influences  significantly the measured voltage in a spin-pumping experiment. Electrical detection of the spin-pumping effect has revealed the crucial role of rectification and shunting effects on the $V_{\mathrm{DC}}$.  The evaluation of the data with the help of a line shape analysis showed  that the rectification generated voltage  and the  ISHE produced voltage have opposite signs. The reduction of the symmetric voltage contribution indicates the increased flow of spin current with decreasing MgO thickness. Future studies on the exact nature of the  flow of spin currents through a few monolayers of insulating layer  could potentially  reveal new phenomena in spin-pumping  structures.

Financial support from Carl Zeiss Stiftung, from the PPP-IKYDA 2015-DAAD bilateral German-Greek Collaboration scheme and from M-era.Net through the HEUMEM project is gratefully acknowledged.


\begin{thebibliography}{26}%
\makeatletter
\providecommand \@ifxundefined [1]{%
 \@ifx{#1\undefined}
}%
\providecommand \@ifnum [1]{%
 \ifnum #1\expandafter \@firstoftwo
 \else \expandafter \@secondoftwo
 \fi
}%
\providecommand \@ifx [1]{%
 \ifx #1\expandafter \@firstoftwo
 \else \expandafter \@secondoftwo
 \fi
}%
\providecommand \natexlab [1]{#1}%
\providecommand \enquote  [1]{``#1''}%
\providecommand \bibnamefont  [1]{#1}%
\providecommand \bibfnamefont [1]{#1}%
\providecommand \citenamefont [1]{#1}%
\providecommand \href@noop [0]{\@secondoftwo}%
\providecommand \href [0]{\begingroup \@sanitize@url \@href}%
\providecommand \@href[1]{\@@startlink{#1}\@@href}%
\providecommand \@@href[1]{\endgroup#1\@@endlink}%
\providecommand \@sanitize@url [0]{\catcode `\\12\catcode `\$12\catcode
  `\&12\catcode `\#12\catcode `\^12\catcode `\_12\catcode `\%12\relax}%
\providecommand \@@startlink[1]{}%
\providecommand \@@endlink[0]{}%
\providecommand \url  [0]{\begingroup\@sanitize@url \@url }%
\providecommand \@url [1]{\endgroup\@href {#1}{\urlprefix }}%
\providecommand \urlprefix  [0]{URL }%
\providecommand \Eprint [0]{\href }%
\providecommand \doibase [0]{http://dx.doi.org/}%
\providecommand \selectlanguage [0]{\@gobble}%
\providecommand \bibinfo  [0]{\@secondoftwo}%
\providecommand \bibfield  [0]{\@secondoftwo}%
\providecommand \translation [1]{[#1]}%
\providecommand \BibitemOpen [0]{}%
\providecommand \bibitemStop [0]{}%
\providecommand \bibitemNoStop [0]{.\EOS\space}%
\providecommand \EOS [0]{\spacefactor3000\relax}%
\providecommand \BibitemShut  [1]{\csname bibitem#1\endcsname}%
\let\auto@bib@innerbib\@empty
\bibitem [{\citenamefont {Hoffmann}(2013)}]{6516040}%
  \BibitemOpen
  \bibfield  {author} {\bibinfo {author} {\bibfnamefont {A.}~\bibnamefont
  {Hoffmann}},\ }\href {\doibase 10.1109/TMAG.2013.2262947} {\bibfield
  {journal} {\bibinfo  {journal} {IEEE Transactions on Magnetics}\ }\textbf
  {\bibinfo {volume} {49}},\ \bibinfo {pages} {5172} (\bibinfo {year}
  {2013})}\BibitemShut {NoStop}%
\bibitem [{\citenamefont {Sinova}\ \emph {et~al.}(2015)\citenamefont {Sinova},
  \citenamefont {Valenzuela}, \citenamefont {Wunderlich}, \citenamefont
  {Back},\ and\ \citenamefont {Jungwirth}}]{RevModPhys.87.1213}%
  \BibitemOpen
  \bibfield  {author} {\bibinfo {author} {\bibfnamefont {J.}~\bibnamefont
  {Sinova}}, \bibinfo {author} {\bibfnamefont {S.~O.}\ \bibnamefont
  {Valenzuela}}, \bibinfo {author} {\bibfnamefont {J.}~\bibnamefont
  {Wunderlich}}, \bibinfo {author} {\bibfnamefont {C.~H.}\ \bibnamefont
  {Back}}, \ and\ \bibinfo {author} {\bibfnamefont {T.}~\bibnamefont
  {Jungwirth}},\ }\href {\doibase 10.1103/RevModPhys.87.1213} {\bibfield
  {journal} {\bibinfo  {journal} {Rev. Mod. Phys.}\ }\textbf {\bibinfo {volume}
  {87}},\ \bibinfo {pages} {1213} (\bibinfo {year} {2015})}\BibitemShut
  {NoStop}%
\bibitem [{\citenamefont {Sun}\ \emph {et~al.}(2013)\citenamefont {Sun},
  \citenamefont {Chang}, \citenamefont {Kabatek}, \citenamefont {Song},
  \citenamefont {Wang}, \citenamefont {Jantz}, \citenamefont {Schneider},
  \citenamefont {Wu}, \citenamefont {Montoya}, \citenamefont {Kardasz},
  \citenamefont {Heinrich}, \citenamefont {te~Velthuis}, \citenamefont
  {Schultheiss},\ and\ \citenamefont {Hoffmann}}]{PhysRevLett.111.106601}%
  \BibitemOpen
  \bibfield  {author} {\bibinfo {author} {\bibfnamefont {Y.}~\bibnamefont
  {Sun}}, \bibinfo {author} {\bibfnamefont {H.}~\bibnamefont {Chang}}, \bibinfo
  {author} {\bibfnamefont {M.}~\bibnamefont {Kabatek}}, \bibinfo {author}
  {\bibfnamefont {Y.-Y.}\ \bibnamefont {Song}}, \bibinfo {author}
  {\bibfnamefont {Z.}~\bibnamefont {Wang}}, \bibinfo {author} {\bibfnamefont
  {M.}~\bibnamefont {Jantz}}, \bibinfo {author} {\bibfnamefont
  {W.}~\bibnamefont {Schneider}}, \bibinfo {author} {\bibfnamefont
  {M.}~\bibnamefont {Wu}}, \bibinfo {author} {\bibfnamefont {E.}~\bibnamefont
  {Montoya}}, \bibinfo {author} {\bibfnamefont {B.}~\bibnamefont {Kardasz}},
  \bibinfo {author} {\bibfnamefont {B.}~\bibnamefont {Heinrich}}, \bibinfo
  {author} {\bibfnamefont {S.~G.~E.}\ \bibnamefont {te~Velthuis}}, \bibinfo
  {author} {\bibfnamefont {H.}~\bibnamefont {Schultheiss}}, \ and\ \bibinfo
  {author} {\bibfnamefont {A.}~\bibnamefont {Hoffmann}},\ }\href {\doibase
  10.1103/PhysRevLett.111.106601} {\bibfield  {journal} {\bibinfo  {journal}
  {Phys. Rev. Lett.}\ }\textbf {\bibinfo {volume} {111}},\ \bibinfo {pages}
  {106601} (\bibinfo {year} {2013})}\BibitemShut {NoStop}%
\bibitem [{\citenamefont {Jungfleisch}\ \emph {et~al.}(2015)\citenamefont
  {Jungfleisch}, \citenamefont {Chumak}, \citenamefont {Kehlberger},
  \citenamefont {Lauer}, \citenamefont {Kim}, \citenamefont {Onbasli},
  \citenamefont {Ross}, \citenamefont {Kl\"aui},\ and\ \citenamefont
  {Hillebrands}}]{Jungfleisch-thickness}%
  \BibitemOpen
  \bibfield  {author} {\bibinfo {author} {\bibfnamefont {M.~B.}\ \bibnamefont
  {Jungfleisch}}, \bibinfo {author} {\bibfnamefont {A.~V.}\ \bibnamefont
  {Chumak}}, \bibinfo {author} {\bibfnamefont {A.}~\bibnamefont {Kehlberger}},
  \bibinfo {author} {\bibfnamefont {V.}~\bibnamefont {Lauer}}, \bibinfo
  {author} {\bibfnamefont {D.~H.}\ \bibnamefont {Kim}}, \bibinfo {author}
  {\bibfnamefont {M.~C.}\ \bibnamefont {Onbasli}}, \bibinfo {author}
  {\bibfnamefont {C.~A.}\ \bibnamefont {Ross}}, \bibinfo {author}
  {\bibfnamefont {M.}~\bibnamefont {Kl\"aui}}, \ and\ \bibinfo {author}
  {\bibfnamefont {B.}~\bibnamefont {Hillebrands}},\ }\href {\doibase
  10.1103/PhysRevB.91.134407} {\bibfield  {journal} {\bibinfo  {journal} {Phys.
  Rev. B}\ }\textbf {\bibinfo {volume} {91}},\ \bibinfo {pages} {134407}
  (\bibinfo {year} {2015})}\BibitemShut {NoStop}%
\bibitem [{\citenamefont {Chen}\ \emph {et~al.}(2013)\citenamefont {Chen},
  \citenamefont {Takahashi}, \citenamefont {Nakayama}, \citenamefont
  {Althammer}, \citenamefont {Goennenwein}, \citenamefont {Saitoh},\ and\
  \citenamefont {Bauer}}]{PhysRevB.87.144411}%
  \BibitemOpen
  \bibfield  {author} {\bibinfo {author} {\bibfnamefont {Y.-T.}\ \bibnamefont
  {Chen}}, \bibinfo {author} {\bibfnamefont {S.}~\bibnamefont {Takahashi}},
  \bibinfo {author} {\bibfnamefont {H.}~\bibnamefont {Nakayama}}, \bibinfo
  {author} {\bibfnamefont {M.}~\bibnamefont {Althammer}}, \bibinfo {author}
  {\bibfnamefont {S.~T.~B.}\ \bibnamefont {Goennenwein}}, \bibinfo {author}
  {\bibfnamefont {E.}~\bibnamefont {Saitoh}}, \ and\ \bibinfo {author}
  {\bibfnamefont {G.~E.~W.}\ \bibnamefont {Bauer}},\ }\href {\doibase
  10.1103/PhysRevB.87.144411} {\bibfield  {journal} {\bibinfo  {journal} {Phys.
  Rev. B}\ }\textbf {\bibinfo {volume} {87}},\ \bibinfo {pages} {144411}
  (\bibinfo {year} {2013})}\BibitemShut {NoStop}%
\bibitem [{\citenamefont {Nakayama}\ \emph {et~al.}(2013)\citenamefont
  {Nakayama}, \citenamefont {Althammer}, \citenamefont {Chen}, \citenamefont
  {Uchida}, \citenamefont {Kajiwara}, \citenamefont {Kikuchi}, \citenamefont
  {Ohtani}, \citenamefont {Gepr\"ags}, \citenamefont {Opel}, \citenamefont
  {Takahashi}, \citenamefont {Gross}, \citenamefont {Bauer}, \citenamefont
  {Goennenwein},\ and\ \citenamefont {Saitoh}}]{PhysRevLett.110.206601}%
  \BibitemOpen
  \bibfield  {author} {\bibinfo {author} {\bibfnamefont {H.}~\bibnamefont
  {Nakayama}}, \bibinfo {author} {\bibfnamefont {M.}~\bibnamefont {Althammer}},
  \bibinfo {author} {\bibfnamefont {Y.-T.}\ \bibnamefont {Chen}}, \bibinfo
  {author} {\bibfnamefont {K.}~\bibnamefont {Uchida}}, \bibinfo {author}
  {\bibfnamefont {Y.}~\bibnamefont {Kajiwara}}, \bibinfo {author}
  {\bibfnamefont {D.}~\bibnamefont {Kikuchi}}, \bibinfo {author} {\bibfnamefont
  {T.}~\bibnamefont {Ohtani}}, \bibinfo {author} {\bibfnamefont
  {S.}~\bibnamefont {Gepr\"ags}}, \bibinfo {author} {\bibfnamefont
  {M.}~\bibnamefont {Opel}}, \bibinfo {author} {\bibfnamefont {S.}~\bibnamefont
  {Takahashi}}, \bibinfo {author} {\bibfnamefont {R.}~\bibnamefont {Gross}},
  \bibinfo {author} {\bibfnamefont {G.~E.~W.}\ \bibnamefont {Bauer}}, \bibinfo
  {author} {\bibfnamefont {S.~T.~B.}\ \bibnamefont {Goennenwein}}, \ and\
  \bibinfo {author} {\bibfnamefont {E.}~\bibnamefont {Saitoh}},\ }\href
  {\doibase 10.1103/PhysRevLett.110.206601} {\bibfield  {journal} {\bibinfo
  {journal} {Phys. Rev. Lett.}\ }\textbf {\bibinfo {volume} {110}},\ \bibinfo
  {pages} {206601} (\bibinfo {year} {2013})}\BibitemShut {NoStop}%
\bibitem [{\citenamefont {Vlietstra}\ \emph {et~al.}(2013)\citenamefont
  {Vlietstra}, \citenamefont {Shan}, \citenamefont {Castel}, \citenamefont {van
  Wees},\ and\ \citenamefont {Ben~Youssef}}]{PhysRevB.87.184421}%
  \BibitemOpen
  \bibfield  {author} {\bibinfo {author} {\bibfnamefont {N.}~\bibnamefont
  {Vlietstra}}, \bibinfo {author} {\bibfnamefont {J.}~\bibnamefont {Shan}},
  \bibinfo {author} {\bibfnamefont {V.}~\bibnamefont {Castel}}, \bibinfo
  {author} {\bibfnamefont {B.~J.}\ \bibnamefont {van Wees}}, \ and\ \bibinfo
  {author} {\bibfnamefont {J.}~\bibnamefont {Ben~Youssef}},\ }\href {\doibase
  10.1103/PhysRevB.87.184421} {\bibfield  {journal} {\bibinfo  {journal} {Phys.
  Rev. B}\ }\textbf {\bibinfo {volume} {87}},\ \bibinfo {pages} {184421}
  (\bibinfo {year} {2013})}\BibitemShut {NoStop}%
\bibitem [{\citenamefont {Ando}\ \emph {et~al.}(2011)\citenamefont {Ando},
  \citenamefont {Takahashi}, \citenamefont {Ieda}, \citenamefont {Kajiwara},
  \citenamefont {Nakayama}, \citenamefont {Yoshino}, \citenamefont {Harii},
  \citenamefont {Fujikawa}, \citenamefont {Matsuo}, \citenamefont {Maekawa},\
  and\ \citenamefont {Saitoh}}]{Ando2011}%
  \BibitemOpen
  \bibfield  {author} {\bibinfo {author} {\bibfnamefont {K.}~\bibnamefont
  {Ando}}, \bibinfo {author} {\bibfnamefont {S.}~\bibnamefont {Takahashi}},
  \bibinfo {author} {\bibfnamefont {J.}~\bibnamefont {Ieda}}, \bibinfo {author}
  {\bibfnamefont {Y.}~\bibnamefont {Kajiwara}}, \bibinfo {author}
  {\bibfnamefont {H.}~\bibnamefont {Nakayama}}, \bibinfo {author}
  {\bibfnamefont {T.}~\bibnamefont {Yoshino}}, \bibinfo {author} {\bibfnamefont
  {K.}~\bibnamefont {Harii}}, \bibinfo {author} {\bibfnamefont
  {Y.}~\bibnamefont {Fujikawa}}, \bibinfo {author} {\bibfnamefont
  {M.}~\bibnamefont {Matsuo}}, \bibinfo {author} {\bibfnamefont
  {S.}~\bibnamefont {Maekawa}}, \ and\ \bibinfo {author} {\bibfnamefont
  {E.}~\bibnamefont {Saitoh}},\ }\href {\doibase 10.1063/1.3587173} {\bibfield
  {journal} {\bibinfo  {journal} {J. Appl. Phys.}\ }\textbf {\bibinfo {volume}
  {109}},\ \bibinfo {pages} {103913} (\bibinfo {year} {2011})}\BibitemShut
  {NoStop}%
\bibitem [{\citenamefont {Vilela-Leao}\ \emph {et~al.}(2011)\citenamefont
  {Vilela-Leao}, \citenamefont {da~Silva}, \citenamefont {Salvador},
  \citenamefont {Rezende},\ and\ \citenamefont {Azevedo}}]{vilela-leao:07C910}%
  \BibitemOpen
  \bibfield  {author} {\bibinfo {author} {\bibfnamefont {L.~H.}\ \bibnamefont
  {Vilela-Leao}}, \bibinfo {author} {\bibfnamefont {G.~L.}\ \bibnamefont
  {da~Silva}}, \bibinfo {author} {\bibfnamefont {C.}~\bibnamefont {Salvador}},
  \bibinfo {author} {\bibfnamefont {S.~M.}\ \bibnamefont {Rezende}}, \ and\
  \bibinfo {author} {\bibfnamefont {A.}~\bibnamefont {Azevedo}},\ }\href
  {\doibase 10.1063/1.3549582} {\bibfield  {journal} {\bibinfo  {journal} {J.
  Appl. Phys.}\ }\textbf {\bibinfo {volume} {109}},\ \bibinfo {eid} {07C910}
  (\bibinfo {year} {2011})}\BibitemShut {NoStop}%
\bibitem [{\citenamefont {Tserkovnyak}\ \emph {et~al.}(2002)\citenamefont
  {Tserkovnyak}, \citenamefont {Brataas},\ and\ \citenamefont
  {Bauer}}]{Tserkovnyak2002}%
  \BibitemOpen
  \bibfield  {author} {\bibinfo {author} {\bibfnamefont {Y.}~\bibnamefont
  {Tserkovnyak}}, \bibinfo {author} {\bibfnamefont {A.}~\bibnamefont
  {Brataas}}, \ and\ \bibinfo {author} {\bibfnamefont {G.}~\bibnamefont
  {Bauer}},\ }\href {\doibase 10.1103/PhysRevLett.88.117601} {\bibfield
  {journal} {\bibinfo  {journal} {Phys. Rev. Lett.}\ }\textbf {\bibinfo
  {volume} {88}},\ \bibinfo {pages} {117601} (\bibinfo {year}
  {2002})}\BibitemShut {NoStop}%
\bibitem [{\citenamefont {Weiler}\ \emph {et~al.}(2013)\citenamefont {Weiler},
  \citenamefont {Althammer}, \citenamefont {Schreier}, \citenamefont {Lotze},
  \citenamefont {Pernpeintner}, \citenamefont {Meyer}, \citenamefont {Huebl},
  \citenamefont {Gross}, \citenamefont {Kamra}, \citenamefont {Xiao},
  \citenamefont {Chen}, \citenamefont {Jiao}, \citenamefont {Bauer},\ and\
  \citenamefont {Goennenwein}}]{PhysRevLett.111.176601}%
  \BibitemOpen
  \bibfield  {author} {\bibinfo {author} {\bibfnamefont {M.}~\bibnamefont
  {Weiler}}, \bibinfo {author} {\bibfnamefont {M.}~\bibnamefont {Althammer}},
  \bibinfo {author} {\bibfnamefont {M.}~\bibnamefont {Schreier}}, \bibinfo
  {author} {\bibfnamefont {J.}~\bibnamefont {Lotze}}, \bibinfo {author}
  {\bibfnamefont {M.}~\bibnamefont {Pernpeintner}}, \bibinfo {author}
  {\bibfnamefont {S.}~\bibnamefont {Meyer}}, \bibinfo {author} {\bibfnamefont
  {H.}~\bibnamefont {Huebl}}, \bibinfo {author} {\bibfnamefont
  {R.}~\bibnamefont {Gross}}, \bibinfo {author} {\bibfnamefont
  {A.}~\bibnamefont {Kamra}}, \bibinfo {author} {\bibfnamefont
  {J.}~\bibnamefont {Xiao}}, \bibinfo {author} {\bibfnamefont {Y.-T.}\
  \bibnamefont {Chen}}, \bibinfo {author} {\bibfnamefont {H.}~\bibnamefont
  {Jiao}}, \bibinfo {author} {\bibfnamefont {G.~E.~W.}\ \bibnamefont {Bauer}},
  \ and\ \bibinfo {author} {\bibfnamefont {S.~T.~B.}\ \bibnamefont
  {Goennenwein}},\ }\href {\doibase 10.1103/PhysRevLett.111.176601} {\bibfield
  {journal} {\bibinfo  {journal} {Phys. Rev. Lett.}\ }\textbf {\bibinfo
  {volume} {111}},\ \bibinfo {pages} {176601} (\bibinfo {year}
  {2013})}\BibitemShut {NoStop}%
\bibitem [{\citenamefont {Czeschka}\ \emph {et~al.}(2011)\citenamefont
  {Czeschka}, \citenamefont {Dreher}, \citenamefont {Brandt}, \citenamefont
  {Weiler}, \citenamefont {Althammer}, \citenamefont {Imort}, \citenamefont
  {Reiss}, \citenamefont {Thomas}, \citenamefont {Schoch}, \citenamefont
  {Limmer}, \citenamefont {Huebl}, \citenamefont {Gross},\ and\ \citenamefont
  {Goennenwein}}]{PhysRevLett.107.046601}%
  \BibitemOpen
  \bibfield  {author} {\bibinfo {author} {\bibfnamefont {F.~D.}\ \bibnamefont
  {Czeschka}}, \bibinfo {author} {\bibfnamefont {L.}~\bibnamefont {Dreher}},
  \bibinfo {author} {\bibfnamefont {M.~S.}\ \bibnamefont {Brandt}}, \bibinfo
  {author} {\bibfnamefont {M.}~\bibnamefont {Weiler}}, \bibinfo {author}
  {\bibfnamefont {M.}~\bibnamefont {Althammer}}, \bibinfo {author}
  {\bibfnamefont {I.-M.}\ \bibnamefont {Imort}}, \bibinfo {author}
  {\bibfnamefont {G.}~\bibnamefont {Reiss}}, \bibinfo {author} {\bibfnamefont
  {A.}~\bibnamefont {Thomas}}, \bibinfo {author} {\bibfnamefont
  {W.}~\bibnamefont {Schoch}}, \bibinfo {author} {\bibfnamefont
  {W.}~\bibnamefont {Limmer}}, \bibinfo {author} {\bibfnamefont
  {H.}~\bibnamefont {Huebl}}, \bibinfo {author} {\bibfnamefont
  {R.}~\bibnamefont {Gross}}, \ and\ \bibinfo {author} {\bibfnamefont
  {S.~T.~B.}\ \bibnamefont {Goennenwein}},\ }\href {\doibase
  10.1103/PhysRevLett.107.046601} {\bibfield  {journal} {\bibinfo  {journal}
  {Phys. Rev. Lett.}\ }\textbf {\bibinfo {volume} {107}},\ \bibinfo {pages}
  {046601} (\bibinfo {year} {2011})}\BibitemShut {NoStop}%
\bibitem [{\citenamefont {Papaioannou}\ \emph {et~al.}(2013)\citenamefont
  {Papaioannou}, \citenamefont {Fuhrmann}, \citenamefont {Jungfleisch},
  \citenamefont {Br{\"a}cher}, \citenamefont {Pirro}, \citenamefont {Lauer},
  \citenamefont {L{\"o}sch},\ and\ \citenamefont
  {Hillebrands}}]{Papaioannou2013}%
  \BibitemOpen
  \bibfield  {author} {\bibinfo {author} {\bibfnamefont {E.~T.}\ \bibnamefont
  {Papaioannou}}, \bibinfo {author} {\bibfnamefont {P.}~\bibnamefont
  {Fuhrmann}}, \bibinfo {author} {\bibfnamefont {M.~B.}\ \bibnamefont
  {Jungfleisch}}, \bibinfo {author} {\bibfnamefont {T.}~\bibnamefont
  {Br{\"a}cher}}, \bibinfo {author} {\bibfnamefont {P.}~\bibnamefont {Pirro}},
  \bibinfo {author} {\bibfnamefont {V.}~\bibnamefont {Lauer}}, \bibinfo
  {author} {\bibfnamefont {J.}~\bibnamefont {L{\"o}sch}}, \ and\ \bibinfo
  {author} {\bibfnamefont {B.}~\bibnamefont {Hillebrands}},\ }\href {\doibase
  10.1063/1.4825167} {\bibfield  {journal} {\bibinfo  {journal} {Appl. Phys.
  Lett.}\ }\textbf {\bibinfo {volume} {103}},\ \bibinfo {pages} {162401}
  (\bibinfo {year} {2013})}\BibitemShut {NoStop}%
\bibitem [{\citenamefont {Zhang}\ \emph {et~al.}(2015)\citenamefont {Zhang},
  \citenamefont {Han}, \citenamefont {Jiang}, \citenamefont {Yang},\ and\
  \citenamefont {Parkin}}]{Parkin2015}%
  \BibitemOpen
  \bibfield  {author} {\bibinfo {author} {\bibfnamefont {W.}~\bibnamefont
  {Zhang}}, \bibinfo {author} {\bibfnamefont {W.}~\bibnamefont {Han}}, \bibinfo
  {author} {\bibfnamefont {X.}~\bibnamefont {Jiang}}, \bibinfo {author}
  {\bibfnamefont {S.-H.}\ \bibnamefont {Yang}}, \ and\ \bibinfo {author}
  {\bibfnamefont {S.~P.}\ \bibnamefont {Parkin}},\ }\href {\doibase
  10.1038/nphys3304} {\bibfield  {journal} {\bibinfo  {journal} {Nat. Phys.}\
  }\textbf {\bibinfo {volume} {11}},\ \bibinfo {pages} {496} (\bibinfo {year}
  {2015})}\BibitemShut {NoStop}%
\bibitem [{\citenamefont {Tserkovnyak}\ \emph {et~al.}(2008)\citenamefont
  {Tserkovnyak}, \citenamefont {Kolodzey}, \citenamefont {Xiao}, \citenamefont
  {Moriyama}, \citenamefont {Cao}, \citenamefont {Fan}, \citenamefont {Xuan},\
  and\ \citenamefont {Nikolic}}]{Moriyama2008}%
  \BibitemOpen
  \bibfield  {author} {\bibinfo {author} {\bibfnamefont {Y.}~\bibnamefont
  {Tserkovnyak}}, \bibinfo {author} {\bibfnamefont {J.}~\bibnamefont
  {Kolodzey}}, \bibinfo {author} {\bibfnamefont {J.~Q.}\ \bibnamefont {Xiao}},
  \bibinfo {author} {\bibfnamefont {T.}~\bibnamefont {Moriyama}}, \bibinfo
  {author} {\bibfnamefont {R.}~\bibnamefont {Cao}}, \bibinfo {author}
  {\bibfnamefont {X.}~\bibnamefont {Fan}}, \bibinfo {author} {\bibfnamefont
  {G.}~\bibnamefont {Xuan}}, \ and\ \bibinfo {author} {\bibfnamefont {B.~K.}\
  \bibnamefont {Nikolic}},\ }\href {\doibase 10.1103/PhysRevLett.100.067602}
  {\bibfield  {journal} {\bibinfo  {journal} {Phys. Rev. Lett.}\ }\textbf
  {\bibinfo {volume} {067602}},\ \bibinfo {pages} {1} (\bibinfo {year}
  {2008})}\BibitemShut {NoStop}%
\bibitem [{\citenamefont {Mosendz}\ \emph {et~al.}(2010)\citenamefont
  {Mosendz}, \citenamefont {Pearson}, \citenamefont {Fradin}, \citenamefont
  {Bader}, \citenamefont {Hoffmann}, \citenamefont {Mosendz}, \citenamefont
  {Pearson}, \citenamefont {Fradin}, \citenamefont {Bader},\ and\ \citenamefont
  {Hoffmann}}]{Mosendz2010}%
  \BibitemOpen
  \bibfield  {author} {\bibinfo {author} {\bibfnamefont {O.}~\bibnamefont
  {Mosendz}}, \bibinfo {author} {\bibfnamefont {J.~E.}\ \bibnamefont
  {Pearson}}, \bibinfo {author} {\bibfnamefont {F.~Y.}\ \bibnamefont {Fradin}},
  \bibinfo {author} {\bibfnamefont {S.~D.}\ \bibnamefont {Bader}}, \bibinfo
  {author} {\bibfnamefont {A.}~\bibnamefont {Hoffmann}}, \bibinfo {author}
  {\bibfnamefont {O.}~\bibnamefont {Mosendz}}, \bibinfo {author} {\bibfnamefont
  {J.~E.}\ \bibnamefont {Pearson}}, \bibinfo {author} {\bibfnamefont {F.~Y.}\
  \bibnamefont {Fradin}}, \bibinfo {author} {\bibfnamefont {S.~D.}\
  \bibnamefont {Bader}}, \ and\ \bibinfo {author} {\bibfnamefont
  {A.}~\bibnamefont {Hoffmann}},\ }\href {\doibase 10.1063/1.3280378}
  {\bibfield  {journal} {\bibinfo  {journal} {Appl. Phys. Lett.}\ }\textbf
  {\bibinfo {volume} {022502}},\ \bibinfo {pages} {96} (\bibinfo {year}
  {2010})}\BibitemShut {NoStop}%
\bibitem [{\citenamefont {Kim}\ \emph {et~al.}(2011)\citenamefont {Kim},
  \citenamefont {Kim},\ and\ \citenamefont {You}}]{Kim2011}%
  \BibitemOpen
  \bibfield  {author} {\bibinfo {author} {\bibfnamefont {D.-H.}\ \bibnamefont
  {Kim}}, \bibinfo {author} {\bibfnamefont {H.-H.}\ \bibnamefont {Kim}}, \ and\
  \bibinfo {author} {\bibfnamefont {C.-Y.}\ \bibnamefont {You}},\ }\href
  {\doibase 10.1063/1.3626593} {\bibfield  {journal} {\bibinfo  {journal}
  {Appl. Phys. Lett.}\ }\textbf {\bibinfo {volume} {99}},\ \bibinfo {pages}
  {072502} (\bibinfo {year} {2011})}\BibitemShut {NoStop}%
\bibitem [{\citenamefont {Du}\ \emph {et~al.}(2013)\citenamefont {Du},
  \citenamefont {Wang}, \citenamefont {Pu}, \citenamefont {Meyer},
  \citenamefont {Woodward}, \citenamefont {Yang},\ and\ \citenamefont
  {Hammel}}]{PhysRevLett.111.247202}%
  \BibitemOpen
  \bibfield  {author} {\bibinfo {author} {\bibfnamefont {C.~H.}\ \bibnamefont
  {Du}}, \bibinfo {author} {\bibfnamefont {H.~L.}\ \bibnamefont {Wang}},
  \bibinfo {author} {\bibfnamefont {Y.}~\bibnamefont {Pu}}, \bibinfo {author}
  {\bibfnamefont {T.~L.}\ \bibnamefont {Meyer}}, \bibinfo {author}
  {\bibfnamefont {P.~M.}\ \bibnamefont {Woodward}}, \bibinfo {author}
  {\bibfnamefont {F.~Y.}\ \bibnamefont {Yang}}, \ and\ \bibinfo {author}
  {\bibfnamefont {P.~C.}\ \bibnamefont {Hammel}},\ }\href {\doibase
  10.1103/PhysRevLett.111.247202} {\bibfield  {journal} {\bibinfo  {journal}
  {Phys. Rev. Lett.}\ }\textbf {\bibinfo {volume} {111}},\ \bibinfo {pages}
  {247202} (\bibinfo {year} {2013})}\BibitemShut {NoStop}%
\bibitem [{\citenamefont {Conca}\ \emph {et~al.}(2016)\citenamefont {Conca},
  \citenamefont {Keller}, \citenamefont {Mihalceanu}, \citenamefont {Kehagias},
  \citenamefont {Dimitrakopulos}, \citenamefont {Hillebrands},\ and\
  \citenamefont {Papaioannou}}]{Conca2016}%
  \BibitemOpen
  \bibfield  {author} {\bibinfo {author} {\bibfnamefont {A.}~\bibnamefont
  {Conca}}, \bibinfo {author} {\bibfnamefont {S.}~\bibnamefont {Keller}},
  \bibinfo {author} {\bibfnamefont {L.}~\bibnamefont {Mihalceanu}}, \bibinfo
  {author} {\bibfnamefont {T.}~\bibnamefont {Kehagias}}, \bibinfo {author}
  {\bibfnamefont {G.~P.}\ \bibnamefont {Dimitrakopulos}}, \bibinfo {author}
  {\bibfnamefont {B.}~\bibnamefont {Hillebrands}}, \ and\ \bibinfo {author}
  {\bibfnamefont {E.~T.}\ \bibnamefont {Papaioannou}},\ }\href {\doibase
  10.1103/PhysRevB.93.134405} {\bibfield  {journal} {\bibinfo  {journal} {Phys.
  Rev. B}\ }\textbf {\bibinfo {volume} {134405}},\ \bibinfo {pages} {1}
  (\bibinfo {year} {2016})}\BibitemShut {NoStop}%
\bibitem [{\citenamefont {Tserkovnyak}\ \emph {et~al.}(2005)\citenamefont
  {Tserkovnyak}, \citenamefont {Brataas}, \citenamefont {Bauer},\ and\
  \citenamefont {Halperin}}]{RevModPhys.77.1375}%
  \BibitemOpen
  \bibfield  {author} {\bibinfo {author} {\bibfnamefont {Y.}~\bibnamefont
  {Tserkovnyak}}, \bibinfo {author} {\bibfnamefont {A.}~\bibnamefont
  {Brataas}}, \bibinfo {author} {\bibfnamefont {G.~E.~W.}\ \bibnamefont
  {Bauer}}, \ and\ \bibinfo {author} {\bibfnamefont {B.~I.}\ \bibnamefont
  {Halperin}},\ }\href {\doibase 10.1103/RevModPhys.77.1375} {\bibfield
  {journal} {\bibinfo  {journal} {Rev. Mod. Phys.}\ }\textbf {\bibinfo {volume}
  {77}},\ \bibinfo {pages} {1375} (\bibinfo {year} {2005})}\BibitemShut
  {NoStop}%
\bibitem [{\citenamefont {Carva}\ and\ \citenamefont
  {Turek}(2007)}]{Carva2007}%
  \BibitemOpen
  \bibfield  {author} {\bibinfo {author} {\bibfnamefont {K.}~\bibnamefont
  {Carva}}\ and\ \bibinfo {author} {\bibfnamefont {I.}~\bibnamefont {Turek}},\
  }\href {\doibase 10.1103/PhysRevB.76.104409} {\bibfield  {journal} {\bibinfo
  {journal} {Phys. Rev. B}\ }\textbf {\bibinfo {volume} {76}},\ \bibinfo
  {pages} {1} (\bibinfo {year} {2007})}\BibitemShut {NoStop}%
\bibitem [{\citenamefont {Juretschke}(1960)}]{Juretschke1960}%
  \BibitemOpen
  \bibfield  {author} {\bibinfo {author} {\bibfnamefont {H.~J.}\ \bibnamefont
  {Juretschke}},\ }\href {\doibase 10.1063/1.1735851} {\bibfield  {journal}
  {\bibinfo  {journal} {J. Appl. Phys.}\ }\textbf {\bibinfo {volume} {31}},\
  \bibinfo {pages} {1401} (\bibinfo {year} {1960})}\BibitemShut {NoStop}%
\bibitem [{\citenamefont {Iguchi}\ and\ \citenamefont
  {Saitoh}(2016)}]{Saitoh2016}%
  \BibitemOpen
  \bibfield  {author} {\bibinfo {author} {\bibfnamefont {R.}~\bibnamefont
  {Iguchi}}\ and\ \bibinfo {author} {\bibfnamefont {E.}~\bibnamefont
  {Saitoh}},\ }\href@noop {} {\  (\bibinfo {year} {2016})},\ \Eprint
  {http://arxiv.org/abs/1607.04716v1} {arXiv:1607.04716v1} \BibitemShut
  {NoStop}%
\bibitem [{\citenamefont {Azevedo}\ \emph {et~al.}(2011)\citenamefont
  {Azevedo}, \citenamefont {Vilela-Le\~{a}o}, \citenamefont
  {Rodr\'{\i}guez-Su\'{a}rez}, \citenamefont {{Lacerda Santos}},\ and\
  \citenamefont {Rezende}}]{Azevedo2011}%
  \BibitemOpen
  \bibfield  {author} {\bibinfo {author} {\bibfnamefont {A.}~\bibnamefont
  {Azevedo}}, \bibinfo {author} {\bibfnamefont {L.~H.}\ \bibnamefont
  {Vilela-Le\~{a}o}}, \bibinfo {author} {\bibfnamefont {R.~L.}\ \bibnamefont
  {Rodr\'{\i}guez-Su\'{a}rez}}, \bibinfo {author} {\bibfnamefont {A.~F.}\
  \bibnamefont {{Lacerda Santos}}}, \ and\ \bibinfo {author} {\bibfnamefont
  {S.~M.}\ \bibnamefont {Rezende}},\ }\href {\doibase
  10.1103/PhysRevB.83.144402} {\bibfield  {journal} {\bibinfo  {journal} {Phys.
  Rev. B}\ }\textbf {\bibinfo {volume} {83}},\ \bibinfo {pages} {144402}
  (\bibinfo {year} {2011})}\BibitemShut {NoStop}%
\bibitem [{\citenamefont {Keller}\ \emph {et~al.}(2017)\citenamefont {Keller},
  \citenamefont {Greser}, \citenamefont {Schweizer}, \citenamefont
  {Hillebrands},\ and\ \citenamefont {Papaioannou}}]{Sascha2017}%
  \BibitemOpen
  \bibfield  {author} {\bibinfo {author} {\bibfnamefont {S.}~\bibnamefont
  {Keller}}, \bibinfo {author} {\bibfnamefont {A.}~\bibnamefont {Greser},
  \bibfnamefont {J~Conca}}, \bibinfo {author} {\bibfnamefont {M.}~\bibnamefont
  {Schweizer}}, \bibinfo {author} {\bibfnamefont {B.}~\bibnamefont
  {Hillebrands}}, \ and\ \bibinfo {author} {\bibfnamefont {E.~T.}\ \bibnamefont
  {Papaioannou}},\ }\href@noop {} {\  (\bibinfo {year} {2017})},\ \Eprint
  {http://arxiv.org/abs/1702.03119} {arXiv:1702.03119} \BibitemShut {NoStop}%
\bibitem [{\citenamefont {Conca}\ \emph {et~al.}(2017)\citenamefont {Conca},
  \citenamefont {Heinz}, \citenamefont {Schweizer}, \citenamefont {Keller},
  \citenamefont {Papaioannou},\ and\ \citenamefont {Hillebrands}}]{Conca2017}%
  \BibitemOpen
  \bibfield  {author} {\bibinfo {author} {\bibfnamefont {A.}~\bibnamefont
  {Conca}}, \bibinfo {author} {\bibfnamefont {B.}~\bibnamefont {Heinz}},
  \bibinfo {author} {\bibfnamefont {M.}~\bibnamefont {Schweizer}}, \bibinfo
  {author} {\bibfnamefont {S.}~\bibnamefont {Keller}}, \bibinfo {author}
  {\bibfnamefont {E.~T.}\ \bibnamefont {Papaioannou}}, \ and\ \bibinfo {author}
  {\bibfnamefont {B.}~\bibnamefont {Hillebrands}},\ }\href@noop {} {\
  (\bibinfo {year} {2017})},\ \Eprint {http://arxiv.org/abs/1701.09110}
  {arXiv:1701.09110} \BibitemShut {NoStop}%
\end{thebibliography}

%

\end{document}